\pdfoutput=1
\documentclass[prl,aps,letterpaper,floatfix,superscriptaddress,nofootinbib,preprintnumbers,twocolumn,10pt]{revtex4-1}

\usepackage{dcolumn}
\usepackage{bm}
\usepackage[dvips]{graphicx}
\usepackage{epsfig}
\usepackage{slashed}
\usepackage{feynmp}
\usepackage{amsmath, amsfonts, amssymb}
\usepackage[normalem]{ulem}
\newcommand{\be}{\begin{equation}}
\newcommand{\ee}{\end{equation}}
\newcommand{\bee}{\begin{equation*}}
\newcommand{\eee}{\end{equation*}}
\newcommand{\bea}{\begin{eqnarray}}
\newcommand{\eea}{\end{eqnarray}}
\newcommand{\bean}{\begin{eqnarray*}}
\newcommand{\eean}{\end{eqnarray*}}

\usepackage{color}

\begin{document}
\preprint{IPPP/16/114,~PITT-PACC-1613,~FERMILAB-PUB-16-510-T,~KCL-PH-TH/2016-61}

\title{Simplified Models for Dark Matter Face their Consistent Completions
}

\author{Dorival Gon\c{c}alves} \email{dorival.goncalves@pitt.edu}
\affiliation{Institute of Particle Physics Phenomenology, Physics Department, Durham University,
Durham DH1 3LE, UK}
\affiliation{Department of Physics and Astronomy, University of Pittsburgh, 3941 O'Hara St., Pittsburgh, PA 15260, USA}

\author{Pedro A. N. Machado} \email{pmachado@fnal.gov}
\affiliation{Instituto de Fisica Teorica IFT-UAM/CSIC, Universidad Autonoma de Madrid
Cantoblanco, 28049 Madrid, Spain}
\affiliation{Theoretical Physics Department, Fermi National Accelerator Laboratory, Batavia, IL, 60510, USA}

\author{Jose Miguel No} \email{j.m.no@sussex.ac.uk}
\affiliation{Department of Physics and Astronomy, University of Sussex, 
Brighton BN1 9QH, UK}
\affiliation{Department of Physics, King's College London, Strand, WC2R 2LS London, UK}


\begin{abstract}
Simplified dark matter models have been recently advocated as a
powerful tool to exploit the complementarity between dark matter
direct detection, indirect detection and LHC experimental
probes. Focusing on pseudoscalar mediators between the dark and
visible sectors, we show that the simplified dark matter model
phenomenology departs significantly from that of consistent
${SU(2)_{\mathrm{L}} \times U(1)_{\mathrm{Y}}}$ gauge invariant
completions. We discuss the key physics simplified models fail to
capture, and its impact on LHC searches. 
Notably, we show that resonant mono-Z searches provide competitive
sensitivities to standard mono-jet analyses at $13$ TeV LHC.
\end{abstract}

\maketitle


\begin{center}
 
\textbf{Introduction}
 
\end{center}

The nature of dark matter (DM) is an outstanding mystery at the 
interface of particle physics and cosmology. At the core of the
current paradigm, is the well motivated
Weakly-Interacting-Massive-Particle (WIMP), a thermal relic in the
GeV-TeV mass range (see~\cite{Bertone:2004pz} for a review). WIMPs may
pertain to a hidden sector, neutral under the Standard Model (SM)
gauge group and interacting with the SM via a
\emph{portal}~\cite{portal}.

The large experimental effort aimed at revealing the nature of DM and
its interactions with the SM proceeds along three main avenues: {\it
  (i)} Low energy direct detection experiments, which measure the
scattering of ambient DM from heavy nuclei. {\it (ii)} Indirect
detection experiments which measure the energetic particles product of
DM annihilations in space. {\it (iii)} DM searches at the Large Hadron
Collider (LHC), where pairs of DM particles could be produced and
manifest themselves as events showing an imbalance in momentum
conservation (through the presence of missing transverse momentum
$E_{T}\hspace{-4mm}/$\hspace{2mm} recoiling against a visible final
state).

The complementarity of different DM search avenues plays a very
important role in the exploration of DM properties, and thus
approaches which allow to fully exploit such complementarity have
received a great deal of attention~\cite{Abdallah:2014hon,Buckley:2014fba,
Matsumoto:2014rxa,Matsumoto:2016hbs,DeSimone:2016fbz}.
The leading two such approaches are effective field theories
(EFTs) and DM simplified models. The latter have increasingly gained
attention as, at the LHC, large missing energy selections render the EFT invalid
for a significant range of the parameter space~\cite{Busoni:2013lha, 
Buchmueller:2013dya, Busoni:2014sya}.

However, it is crucial that the simplified models do correctly
describe the relevant physics that a realistic theory beyond the SM
would yield at the LHC, direct and indirect detection
experiments, at least in some limit.  In this Letter, we show
that for simplified DM models with a pseudoscalar mediator, the
minimal consistent $SU(2)_{\mathrm{L}} \times U(1)_{\mathrm{Y}}$ gauge
invariant completions to which these simplified models may be mapped
yield a very different physical picture, signaling a failure of the
simplified models to capture part of the key DM physics: these models
are ``over-simplified" (see \cite{Kahlhoefer:2015bea, Englert:2016joy,
  Duerr:2016tmh, Boveia:2016mrp, Bauer:2016gys} for recent discussions
on this issue). We detail the physics that such simplified models are
neglecting, and show that it has a critical impact on DM searches at the LHC.
Particularly, we demonstrate that the resonant mono-$Z$ channel displays 
competitive sensitivities to the usual mono-jet analysis at the LHC Run-II.

\begin{center}
 
\textbf{Simplified Pseudoscalar Portal for Dark Matter}
 
\end{center}

The simplified model DM scenario that we consider consists of a gauge singlet DM 
fermion $\chi$ (for concreteness we assume a Dirac fermion,
our results can be easily generalised to Majorana fermions~\cite{DeSimone:2016fbz}), whose interactions with the SM occur via a pseudoscalar
mediator $a$~\cite{Buckley:2014fba,Haisch:2012kf,Fox:2012ru,Harris:2014hga,Haisch:2013fla,Mattelaer:2015haa},
namely
\begin{eqnarray}
\label{simplified_model}
\mathcal{L}_{s} &=& \bar{\chi} (i \partial \hspace{-2mm}\slash\hspace{2mm} - m_{\chi})\chi + \frac{1}{2} (\partial_{\mu} a)^2 -\frac{m_{a}^2}{2} a^2 \nonumber \\
&-&  g_{\chi} \,a\, \bar{\chi}i
\gamma^5\chi - g_{\mathrm{SM}} \,a  \sum_{f} \frac{y_{f}}{\sqrt{2}}\, \bar{f}i
\gamma^5f\, .
\end{eqnarray}
We emphasize that a built-in assumption in these scenarios is that DM
belongs to a hidden sector, neutral under SM gauge
interactions\footnote{Departing from this assumption would
dramatically modify DM phenomenology at the LHC, direct and
indirect DM detection experiments.}. The model
in Eq.~\eqref{simplified_model} is described in terms of four parameters:
the masses for the DM $m_{\chi}$ and the mediator $m_{a}$, and the
couplings of the mediator to DM $g_{\chi}$ and the SM\footnote{The
models assume a Minimal Flavour Violation scenario.
A universal rescaling $g_{\mathrm{SM}}$ can be generalized
within the simplified model framework.} $g_{\mathrm{SM}}$.
Assuming the DM candidate $\chi$ to be a thermal relic which obtains
its abundance via freeze-out, the relevant early Universe annihilation
channels for $\chi$ are into bottom quarks (for $m_{\chi} > m_b$), top
quarks (for $m_{\chi} > m_t$) and mediators (for $m_{\chi} > m_{a}$).
The respective thermally averaged annihilation cross sections are
\begin{eqnarray}
\left\langle \sigma \, \mathrm{v} \right\rangle_{\bar{f}f}& = & \frac{3 \,g_{\chi}^2\,g_{\mathrm{SM}}^2\,m_f^2}{4 \pi v^2} \, 
\frac{m_{\chi}^2 \, \sqrt{1- \frac{m_f^2}{m_{\chi}^2}}}{(m_a^2-4m_{\chi}^2)^2 + m_a^2 \Gamma_a^2}\, ,\nonumber\\
\left\langle \sigma \, \mathrm{v} \right\rangle_{aa}& = & \frac{g_{\chi}^4\left(1-\frac{m_a^2}{m_\chi^2}\right)^{3/2}}{8m_\chi^2\left(2-\frac{m_a^2}{m_\chi^2}\right)^2} \,,
\end{eqnarray}
where $v = 246$ GeV. We note that the observed DM relic abundance is obtained for 
$\langle\sigma \mathrm{v}\rangle  \simeq 3\cdot 10^{-26}{\rm  cm}^3/{\rm s}$.

Concerning DM direct detection, the pseudoscalar portal yields a spin-dependent and spin-independent cross section respectively at tree-level and one-loop. 
The experimental constraints are overall found to be extremely weak~\cite{Ipek:2014gua}, and so we can safely disregard DM direct detection
in the following discussion. We also postpone a detailed discussion of indirect detection constraints for the future~\cite{GMN}, and 
focus in this work on DM relic density and collider searches.

\begin{center}
 
\textbf{Gauge-Invariant~Models:~Pseudoscalar~Mediator}
 
\end{center}

A consistent realization of the simplified model scenario displayed in Eq.~\eqref{simplified_model}, respecting
$SU(2)_{\mathrm{L}} \times U(1)_{\mathrm{Y}}$ gauge invariance, 
is obtained along two possible avenues: 

\vspace{1mm}

\noindent
\textsl{(i)} Extending the SM scalar sector with a field that couples to SM fermions and yields pseudoscalar mixing 
with the real mediator field $a$.

\vspace{1mm}

\noindent
\textsl{(ii)} Allowing the SM fermions to mix with heavy vector-like fermion partners
   $\psi$, which couple to the pseudoscalar mediator $a$ via  $g_a a\bar\psi i\gamma^5\psi$.

\vspace{1mm}

In the latter case, the couplings between $a$ and the SM fermions 
are weighted by the Yukawa couplings $y_f$ as in the simplified model Eq.~\eqref{simplified_model}, and the 
$g_{{\rm SM},i}$ parameter (one for each SM fermion) is related to the product of the fermion mixing 
and $g_a$. In this scenario the top/bottom partner mixing plays the most important role, and 
the model needs to incorporate a custodial symmetry to comply with constraints from electroweak precision observables, 
particularly the $Zbb$ coupling and the $T$ parameter. 
The phenomenology of such scenarios will be studied in a future manuscript~\cite{GMN}. 

In this work we analyze in detail the former scenario, hereinafter referred to as \textsl{pseudoscalar portal mixing} scenario.
The dark sector Lagrangian, in terms of the DM $\chi$ and pseudoscalar mediator $a_0$, both $SU(2)_{\mathrm{L}} \times U(1)_{\mathrm{Y}}$ singlets, 
is simply written as
\begin{equation}
\label{Ldark}
 V_{\mathrm{dark}} = \frac{m^2_{a_0}}{2}\,a_0^2 + m_{\chi}\, \bar{\chi}\chi + g_{\chi}\,a_0 \,\bar{\chi} i\gamma^{5} \chi\, .
\end{equation}
We extend the SM Higgs sector to include two scalar doublets $H_{1,2}$~\cite{Nomura:2008ru}. The scalar potential for the two
Higgs doublets, assuming CP-conservation and a softly broken $\mathbb{Z}_2$ symmetry, reads
\begin{eqnarray}	
\label{2HDM_potential}
V_{\mathrm{2HDM}} &= &\mu^2_1 \left|H_1\right|^2 + \mu^2_2\left|H_2\right|^2 - \mu^2\left[H_1^{\dagger}H_2+\mathrm{h.c.}\right] \nonumber \\
&+&\frac{\lambda_1}{2}\left|H_1\right|^4 +\frac{\lambda_2}{2}\left|H_2\right|^4 + \lambda_3 \left|H_1\right|^2\left|H_2\right|^2 \\
&+&\lambda_4 \left|H_1^{\dagger}H_2\right|^2+ \frac{\lambda_5}{2}\left[\left(H_1^{\dagger}H_2\right)^2+\mathrm{h.c.}\right]\, ,\nonumber 
\end{eqnarray}
and the portal between visible and hidden sectors occurs 
via $V_{\mathrm{portal}} = i\,\kappa\,a_0 \,H_1^{\dagger}H_2 + \mathrm{h.c.}$~\cite{Nomura:2008ru,Ipek:2014gua,No:2015xqa}.
The two doublets are $H_i = \left(\phi_i^{+} , (v_i + h_i + \eta_i)/\sqrt{2} \right)^T$, with $v_i$ their \textit{vev} 
($\sqrt{v^2_1 + v^2_2} = v $, $v_2/v_1 \equiv \mathrm{tan} \beta = t_{\beta}$).
The scalar spectrum contains a charged scalar $H^{\pm} = c_\beta \,\phi_2^{\pm} - s_\beta \, \phi_1^{\pm}$
($c_{\beta} \equiv \mathrm{cos}\beta$, $s_{\beta} \equiv \mathrm{sin}\beta$)
and two neutral CP-even scalars $h = c_\alpha \,h_2 - s_\alpha \, h_1$, $H = - s_\alpha \,h_2 - c_\alpha \, h_1$, with
$h$ identified as the 125 GeV Higgs state (SM-like in the alignment limit $\beta - \alpha = \pi/2$~\cite{Gunion:2002zf}). 
The neutral CP-odd scalar $A_0 = c_\beta \,\eta_2 - s_\beta \, \eta_1$ mixes 
with $a_0$ for $\kappa \neq 0$, yielding two pseudoscalar mass eigenstates $a,A$ (with $m_A > m_a$): 
$A = c_{\theta} \,A_0 + s_{\theta} \, a_0$, $a = c_{\theta} \,a_0 - s_{\theta} \, A_0$. 
In terms of the mass eigenstates, we get 
\begin{eqnarray}	
\label{Vportal_mass}
V_{\mathrm{dark}} &\supset& y_{\chi}\,(c_{\theta} \,a + s_{\theta} \, A) \, \bar{\chi} i\gamma^{5} \chi \, , \nonumber \\
V_{\mathrm{portal}} &=& \frac{\left( m_A^2 - m_a^2 \right) s_{2\theta}}{2\,v} \, \left(c_{\beta-\alpha}\,H - s_{\beta-\alpha}\,h \right)  \\
&\times& \left[ a A \,(s_{\theta}^2 - c_{\theta}^2) + (a^2 - A^2)\, s_{\theta} c_{\theta}  \right] \, . \nonumber 
\end{eqnarray}
The coupling of the pseudoscalar mediators $a, A$ to the SM fermions
occurs via the Yukawa couplings of the scalar doublets $H_{1,2}$.  We
consider a scenario with all SM fermions coupled to the same doublet
(2HDM Type I), and another with down-type and up-type quarks coupled
to different doublets (2HDM Type II) (see {\it
e.g.}~\cite{Branco:2011iw} for details).  In the first case, the
couplings of $a$ $(A)$ to SM fermions are all weighted by
$s_\theta\,t_{\beta}^{-1}$ ($c_\theta\, t_{\beta}^{-1}$). In the
second scenario, the weight is $s_{\theta}\,t_{\beta}^{-1}$
($c_\theta\, t_{\beta}^{-1}$) for up-type quarks and
$s_\theta\,t_\beta$ ($c_\theta\,t_\beta$) for down-type quarks. 
We note that for Type II the alignment limit $c_{\beta-\alpha} = 0$ is favoured~\cite{Aad:2015pla}, and thus 
this scenario will be considered in the rest of this letter.
The new scalars also impact electroweak precision observables~\cite{Grimus:2007if},
and we fix in the following $m_{H^{\pm}} \simeq m_{H}$ to satisfy $T$-parameter
bounds~\cite{Gerard:2007kn}.

\vspace{1mm}

\begin{figure}[t!]
\centering
  \includegraphics[width=.48\textwidth]{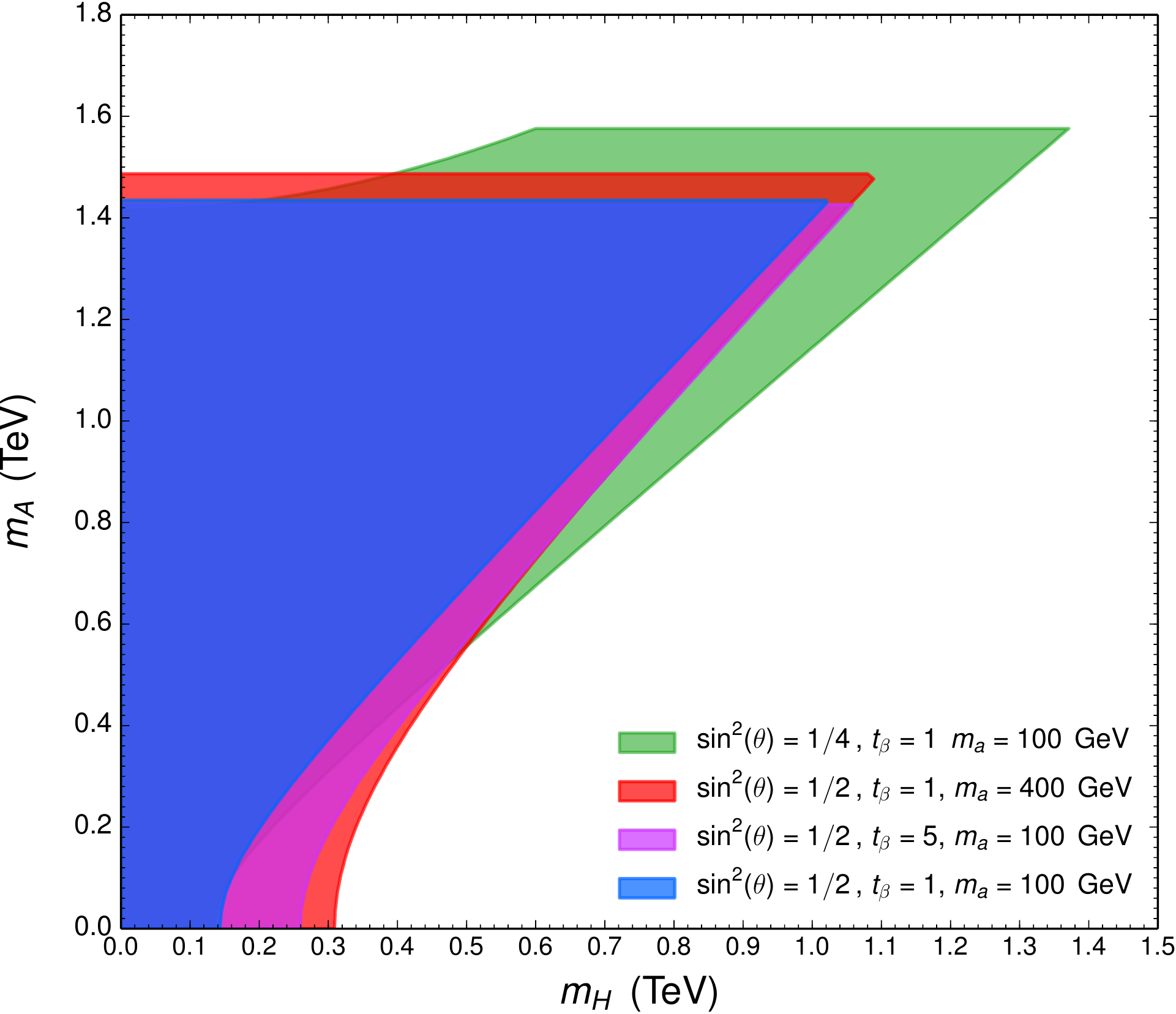}
  
  \vspace{-3mm}  
  
\caption{Allowed parameter space in the ($m_{H},\,m_A$) from unitarity
  and stability constraints (see text for details).  }

\vspace{-2mm}

\label{Fig_Unitarity} 
\end{figure}

We now confront the pseudoscalar portal mixing scenario with the simplified
model pseudoscalar portal in Eq.~\eqref{simplified_model}. First, we note that the portal
interaction can be rewritten as
\begin{equation}
\label{portal}
\kappa = \frac{m_A^2-m_a^2}{2v}\,s_{2\theta} \,.
\end{equation}
Thus, in the presence of mixing $s_{2\theta} \neq 0$,  the portal
interaction grows larger as the mass of $A$ increases.  The unitarity of
scattering processes $a a, A A, a A \to W^{+} W^{-}$ yields an upper
bound on $\Delta_a^2 = m_A^2 - m_a^2$, leading to a non-decoupling of
the states $A$, $H$ and $H^{\pm}$.  We compute the scattering
amplitude matrix $\mathcal{M}_{i\,j\to WW}$ ($i,j = a,A$), choosing
$c_{\beta-\alpha} = 0$ as a conservative assumption since the
unitarity bounds are stronger away from alignment. The amplitudes read
in this limit
\begin{eqnarray}
\mathcal{M}_{AA\to WW} & =& \frac{g^2}{2} c_{\theta}^2 \left(\frac{\Delta_H^2}{m_{W}^2} + \frac{\Delta_a^2 (1-c_{\theta}^2)}{2 m_{W}^2} \right) \, ,  \\
\mathcal{M}_{Aa\to WW} &=& - \frac{g^2}{2} s_{\theta} c_{\theta} \left(\frac{\Delta_H^2}{m_{W}^2} - \frac{\Delta_a^2 (\frac{1}{2}-s_{\theta}^2)}{2 m_{W}^2} 
\right) \, , \\
\mathcal{M}_{aa\to WW} &=& \frac{g^2}{2} s_{\theta}^2 \left(\frac{\Delta_H^2}{m_{W}^2} - \frac{3 \Delta_a^2 c_{\theta}^2}{2 m_{W}^2} \right) \, ,
\end{eqnarray}
with $\Delta_H^2 = M^2 - m_{H^{\pm}}^2 + 2 m_W^2 - m_h^2/2$, $M^2
\equiv \mu^2/(s_{\beta}c_{\beta})$,  neglecting
$\mathcal{O}(1/t,1/s)$ terms.  The eigenvalues of the scattering
amplitude matrix are given by
\begin{eqnarray}
\Lambda_{\pm} = \left[\frac{\Delta_H^2}{v^2}  - \frac{\Delta_a^2 \,(1-c_{4\theta})}{8\,v^2} 
\pm \sqrt{\frac{\Delta_H^4}{v^4}  + \frac{\Delta_a^4 \,(1-c_{4\theta})}{8\,v^4}} \right] , \,\,\,\,\,\,\,\,\,
\end{eqnarray}
and the unitarity bound on the scattering processes $a a, A A, a A \to
W^{+} W^{-}$ is given by $|\Lambda_{\pm}| \leq 8 \pi$.  In addition, a
set of unitarity bounds restrict the values of the quartic
interactions in Eq.~\eqref{2HDM_potential}~\cite{Kanemura:1993hm,
  Akeroyd:2000wc, Ginzburg:2005dt, Grinstein:2015rtl}, which in
combination with boundness from below conditions on the scalar
potential limits the allowed parameter ranges
(see {\it e.g.} the discussion
in~\cite{Kling:2016opi,Dorsch:2016tab}).  The combination of bounds
yields an allowed region in the ($m_{H},\,m_A$) mass plane, weakly
dependent on $m_a$ and $t_{\beta}$, as shown in
Fig.~\ref{Fig_Unitarity}.  While the allowed region increases as
$s_{\theta}$ decreases, the states $A$, $H^{\pm}$, $H$
cannot be heavier than $\mathcal{O}$(TeV) if the portal is active.

The DM thermal relic abundance provides a remarkable piece of information in the above context. 
In order not to overclose the universe a minimum value of the coupling $g_{\rm SM}$ between $a$
and the SM fermions is required. This yields a minimum value of the mixing $s_\theta$ for a
fixed $t_{\beta}$ value\footnote{We recall that for Type I (Type II) $g_{\rm SM} = s_{\theta} t_{\beta}^{-1}$ ($g_{\rm SM} = s_{\theta} t_{\beta}^{-1}$ for $t$ quarks and 
$g_{\rm SM} = s_{\theta} t_{\beta}$ for $b$ quarks).}. 
At the same time, charged scalar loop contributions to the
$\bar B\to X_s \gamma$ flavour process~\cite{Hermann:2012fc,
Misiak:2015xwa} on the ($m_{H^{\pm}}, \, t_{\beta}$) plane yield a lower limit on $t_{\beta}$. 
Since, as discussed previously,
$H^{\pm}$ cannot be heavier than $\mathcal{O}$(TeV) from unitarity if 
$s_{\theta} \neq 0$ (as needed from relic density considerations), the requirement
$m_{H^{\pm}} < 1$ TeV results in the 
bound $t_{\beta} \gtrsim 0.8$ for 2HDM Type I, which then translates into an upper bound on $g_{\rm SM}$. 
Similar upper (lower) bounds on $g_{\rm SM}$ from the lower $\bar B\to X_s \gamma$ 
bound on $t_{\beta}$ apply for 2HDM Type II when the DM annihilates 
dominantly into top (bottom) quarks.

\begin{figure}[h]
\centering

\hspace{-5mm}\includegraphics[width=.48\textwidth]{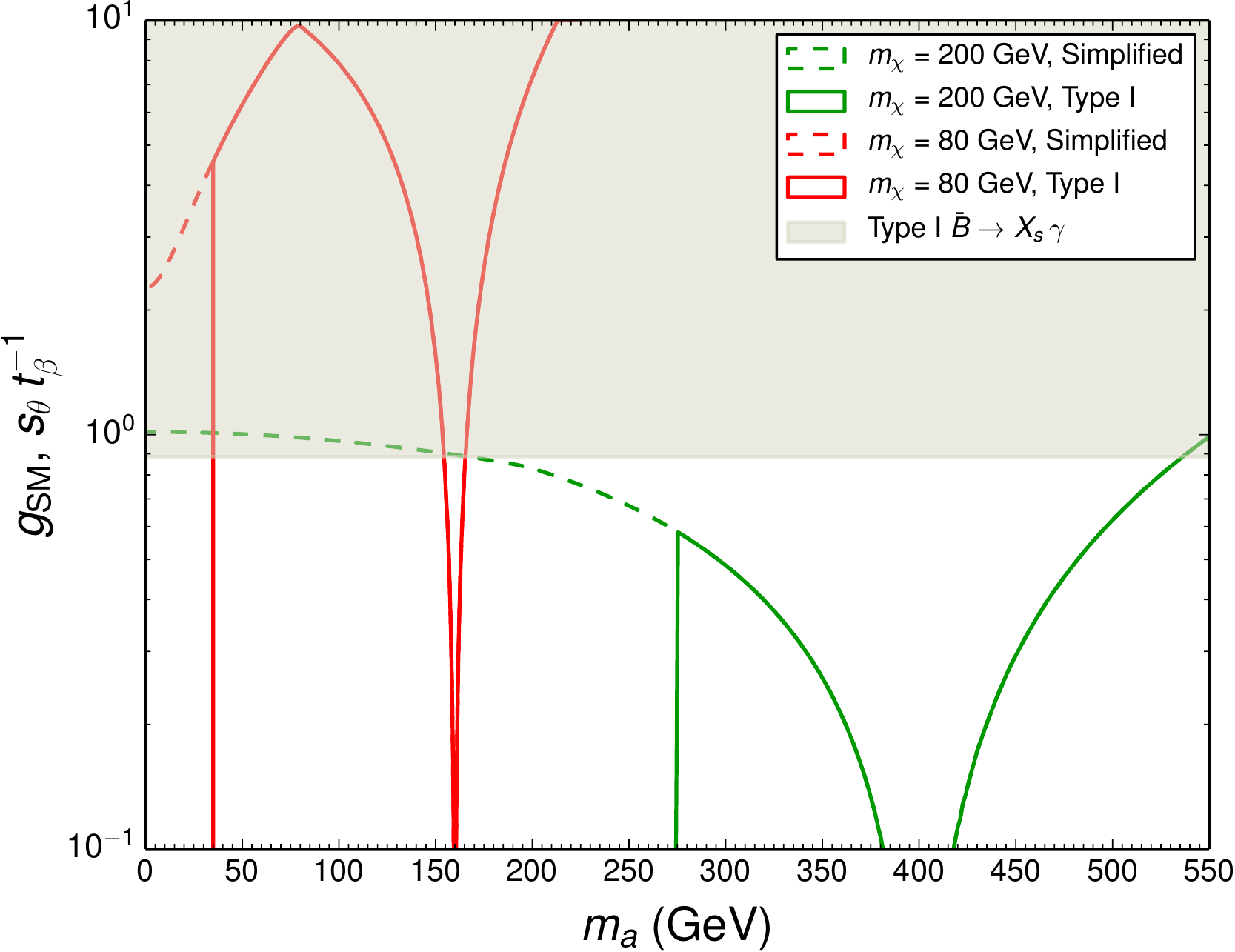}

\vspace{2mm}

\includegraphics[width=.23\textwidth]{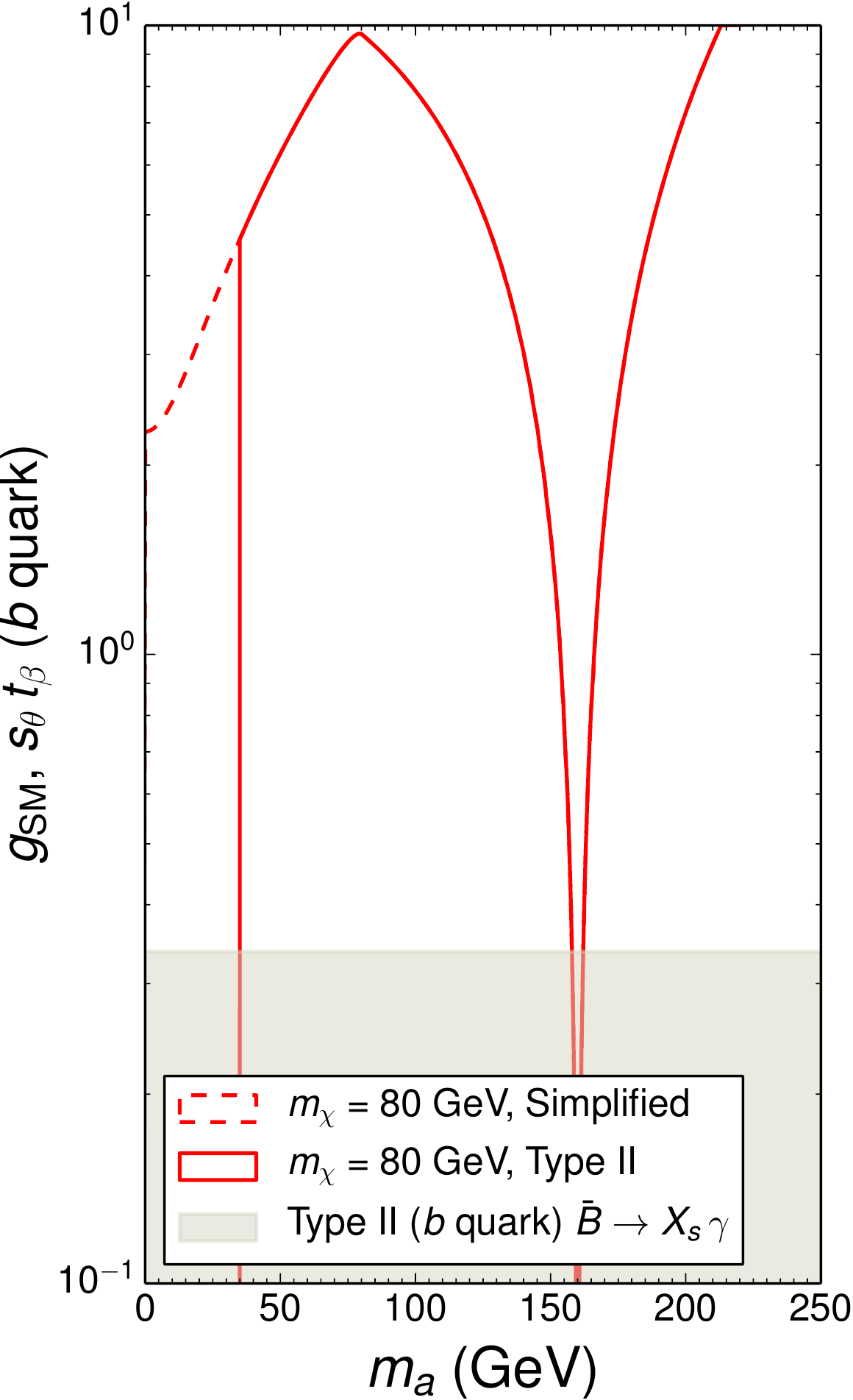}
\includegraphics[width=.23\textwidth]{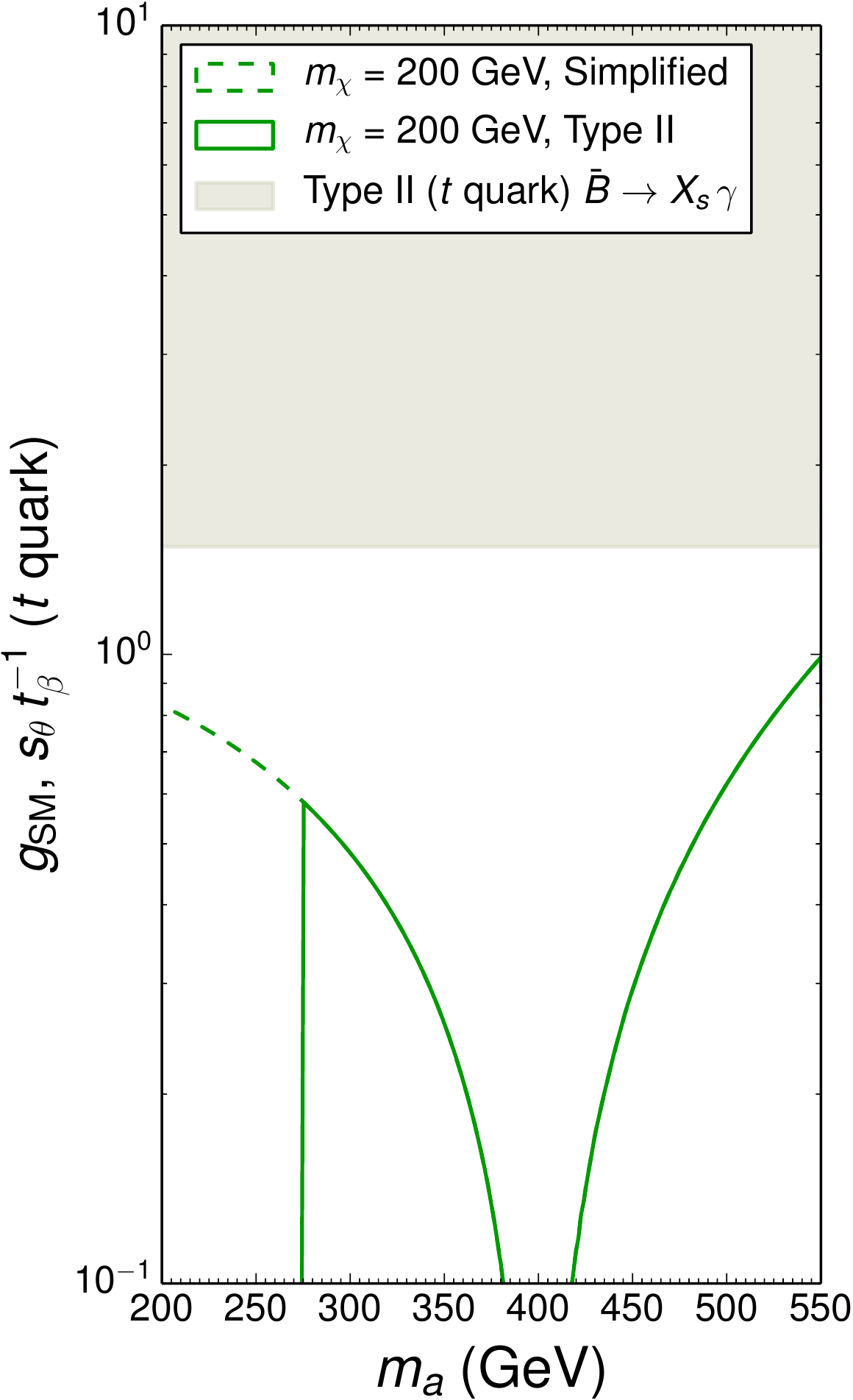}
\vspace{-2mm}
\caption{Relic density comparison between simplified model and 2HDM Type I (upper panel) and Type II 
(lower panels) completion for $\mathrm{sin}^2(\theta) = 1/2$, $m_{H^\pm}=m_{H}=1$~TeV, $m_A=1.4$ TeV, $g_{\chi} = 0.15$, $c_{\beta-\alpha} = 0$.}
\label{Fig_Relic_Pseudo} 
\end{figure}

Besides these important constraints on $g_{\rm SM}$ which are not present in the simplified model, 
another key difference between the consistent completion and the simplified 
model is the presence of new DM annihilation channels $\bar{\chi}\chi \to a\,h, Z\,h$ (the latter for
$c_{\beta-\alpha} \neq 0$), which can be the dominant DM annihilation process for 
heavy DM and light $a$. 
Particularly, the annihilation into $a\, h$ is maximal in the alignment limit $c_{\beta-\alpha} = 0$, for which 
the cross section reads
\begin{eqnarray}
\left\langle \sigma \, \mathrm{v} \right\rangle_{ah}& = & \frac{g_{\chi}^2\,s_{2\theta}^2}{64 \pi^2 \,v^2} \, 
\sqrt{1- \frac{(m_a + m_h)^2}{4 m_{\chi}^2}} (m_A^2- m_a^2)^2 \, \nonumber \\
& \times & \left(\frac{c_\theta s_{2\theta}}{2 (m_a^2 - 4 m_{\chi}^2)} -  \frac{s_\theta c_{2\theta} }{(m_A^2 - 4 m_{\chi}^2)}  \right)^2 \, .
\end{eqnarray}
The relic density comparison between simplified model and consistent completion discussed above
is illustrated in Fig.~\ref{Fig_Relic_Pseudo}, where the value of $g_{\rm SM}$ required to yield 
$\langle\sigma \mathrm{v}\rangle \simeq 3\cdot 10^{-26}{\rm cm}^3/{\rm s}$ 
for simplified model and consistent completion is shown respectively in dashed and solid lines, for 2HDM Type I (Fig.~\ref{Fig_Relic_Pseudo} upper) and 
Type II (Fig.~\ref{Fig_Relic_Pseudo} lower) in the $(m_a, g_{\rm SM})$ plane.
In each case, we consider as illustration $m_{\chi} = 80$ GeV, below the $\bar{t} t$ annihilation
threshold  with $\bar{\chi} \chi \to \bar{b} b$ becoming important,
and $m_{\chi} = 200$ GeV, above the $\bar{t} t$ annihilation threshold. 
To understand the features of these curves, consider {\it e.g.} the $m_{\chi} = 80$ GeV scenario for Type II 2HDM (Fig.~\ref{Fig_Relic_Pseudo}, bottom-left). 
When $m_a > 2 m_\chi = 160$ GeV, the value of $g_{\rm SM}$ required to yield the relic abundance annihilation cross section 
is quite large. As $m_a \to 2m_\chi$, the $\bar{\chi} \chi \to \bar{b} b$ process becomes resonant (modulated by the narrow width of $a$) 
resulting in a much smaller value of $g_{\rm SM}$. For $m_a < m_\chi$, the $t$-channel annihilation process $\bar{\chi}\chi \to a\,a$ opens 
up, and the required value of $g_{\rm SM}$ decreases again. 
Finally, in the consistent completion the annihilation channel $\bar{\chi}\chi \to a\,h$ becomes avaliable for 
$m_{\chi} > (m_a + m_h)/2$, which in this case leads to a depletion of the relic abundance since 
$\left\langle \sigma \, \mathrm{v} \right\rangle_{ah} >  3\cdot 10^{-26}{\rm cm}^3/{\rm s}$ (regardless of the value of $g_{\rm SM}$),
yielding the sharp kink observed in the solid-red line.

\pagebreak

\vspace{-2mm}

As highlighted in Fig~\ref{Fig_Relic_Pseudo}, for 2HDM Type I and $m_{\chi} < m_t$, the $t_{\beta}$ flavour 
bound constrains the consistent completion to the resonant $\bar{\chi}\chi\to \bar{b}b$
annihilation region, or to the region $m_{\chi} \gtrsim m_a$ where the
new annihilation channels $\bar{\chi}\chi \to a\,h, Z\,h$ may be important. 
Above the top-quark
threshold, simplified model and consistent completion yield the same
result both for 2HDM Type I and II, except again for $m_{\chi} > (m_a + m_h)/2$ and/or
$c_{\beta-\alpha} \neq 0$ (with $\bar{\chi}\chi \to Z\,h$ open) where
the new channels play a key role. 
It also follows from this discussion
that indirect DM detection in the gauge-invariant completion and the
simplified model may differ significantly~\cite{GMN}.  However, we
show in the following that it is in the context of LHC searches where
the difference between simplified model and consistent completion
becomes crucial.

\vspace{2mm}

\begin{figure}[t!]
\centering
  \includegraphics[width=.4\textwidth]{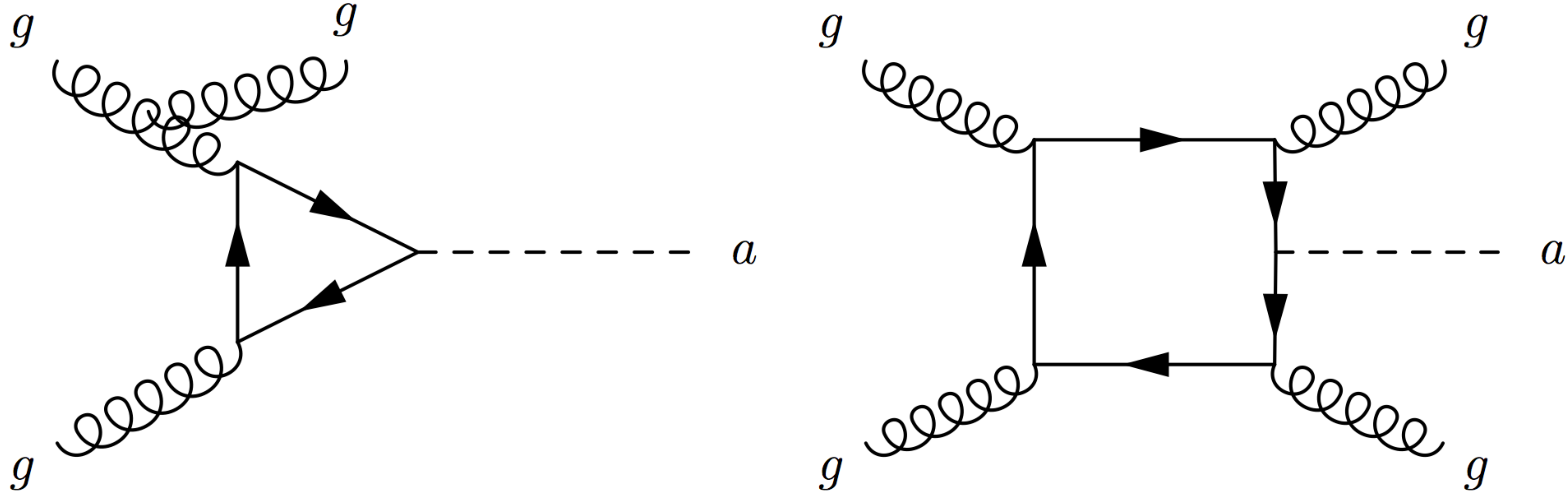}
  \vspace{-1mm}
  \caption{Feynman diagrams for ${pp\rightarrow a}+$jets production with up to two extra jets. 
}
\label{fig:feyn} 
\end{figure}

\noindent \textbf{LHC Phenomenology: Mono-Jet Searches}.~We first study the 
collider phenomenology of  the pseudoscalar resonances for ``mono-jets" 
searches, ${pp\rightarrow a+}$jets. The canonical signal is defined by the presence of large missing
energy, from the pseudoscalar decay to DM, $a\rightarrow \bar{\chi}\chi$, recoiling against one or more 
jets. A sample of Feynman diagrams contributing to the signal is shown in Fig.~\ref{fig:feyn}.

For our analysis, we generate the signal sample with \textsc{Sherpa+OpenLoops}~\cite{sherpa,openloops} 
merging up to two extra jets via the~\textsc{CKKW} algorithm~\cite{ckkw} and accounting for the heavy quark mass effects to the 
pseudoscalar production. Notably,  these mass effects result in relevant changes to the $\slashed{E}_T$ distribution above
$m_t$~\cite{Buckley:2014fba,Buschmann:2014sia,Corbett:2015ksa}, precisely the most sensitive region for the mono-jet search. 
We include NLO QCD corrections through the scaling factor ${K\sim 1.5}$~\cite{Buschmann:2014sia}. 
Hadronization and underlying event effects are also simulated.

\begin{figure}[b!]
\centering
  \includegraphics[width=.48\textwidth]{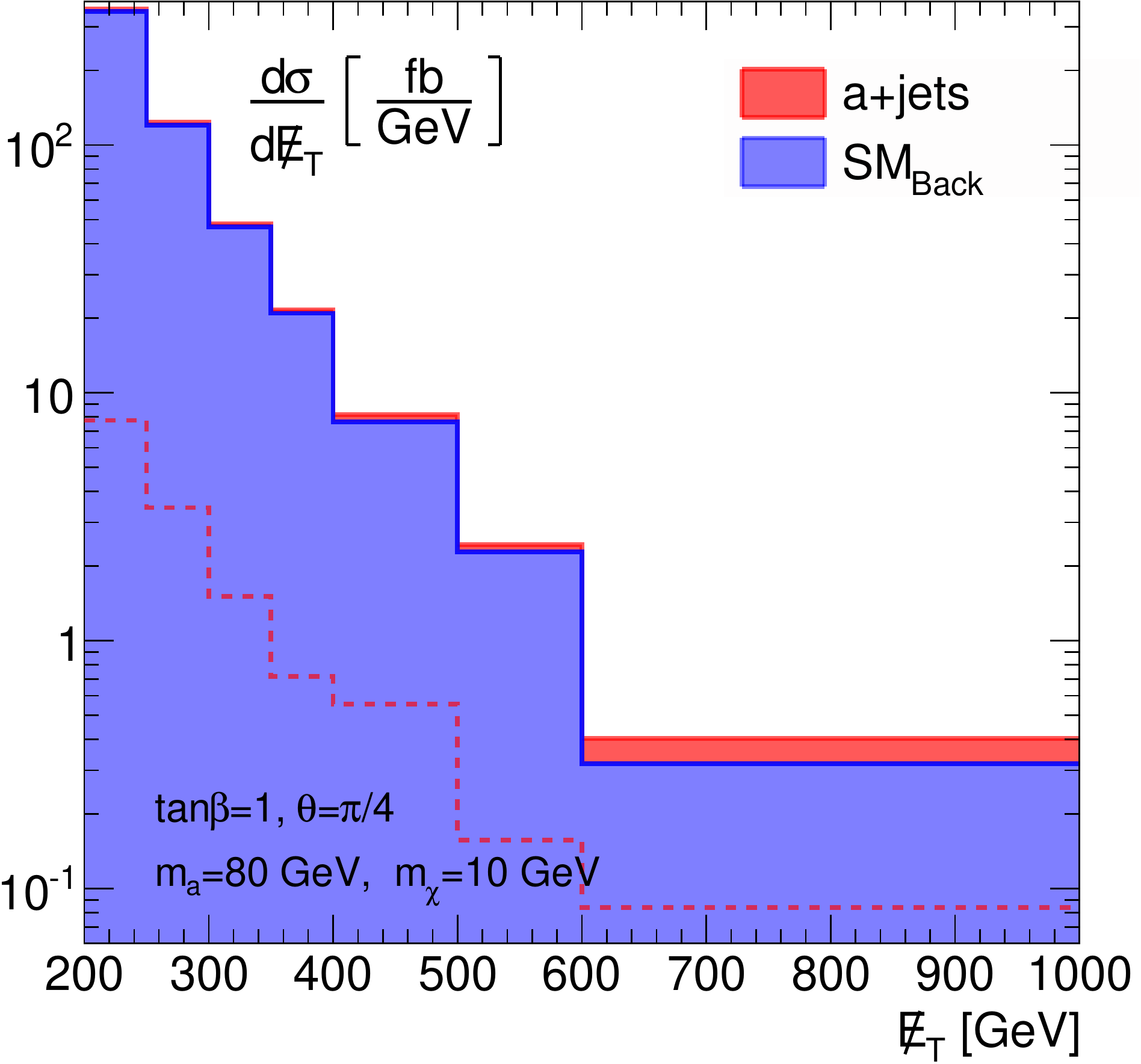}
 
  \vspace{-1mm}

  \caption{Signal (red) and background (blue) transverse missing energy $\slashed{E}_T$
    distributions. Shaded (empty) histograms are (non-)stacked. We assume $m_\chi=10$~GeV, 
    $\mathrm{sin}^2\theta=1/2$, $\tan\beta=g_{\chi}=1$ and $m_a=80$~GeV (with $m_A \gg
    m_a$).  The SM background was obtained from from~\cite{cms_monojet}.}
\label{fig:ptmiss1} 
\end{figure}

Following the recent 13~TeV \textsc{CMS} $\slashed{E}_T +$ jets analysis~\cite{cms_monojet}, we define jets with the anti-$k_T$ 
algorithm $R=0.4$, $p^j_{T}>30$~GeV and $|\eta_j|<2.5$ via \textsc{FastJet}~\cite{fastjet}. b-jets are vetoed with 70\% b-tagging 
efficiency and 1\% mistag rate~\cite{btagging}. Electrons and muons with  $p^{\ell}_{T}>10$~GeV and $|\eta_{\ell}|<2.5$ are rejected. 
To suppress  the $Z$+jets background, events are selected with $p^{j_1}_{T}>100$~GeV for the leading jet and $\slashed{E}_T>200$~GeV. 
Finally, to further reduce the multi-jet background, the azimuthal angle between the $\slashed{E}_T$ direction and the first four 
leading jets is required to be $> 0.5$.

\begin{figure}[t!]
\centering
\includegraphics[width=.48\textwidth]{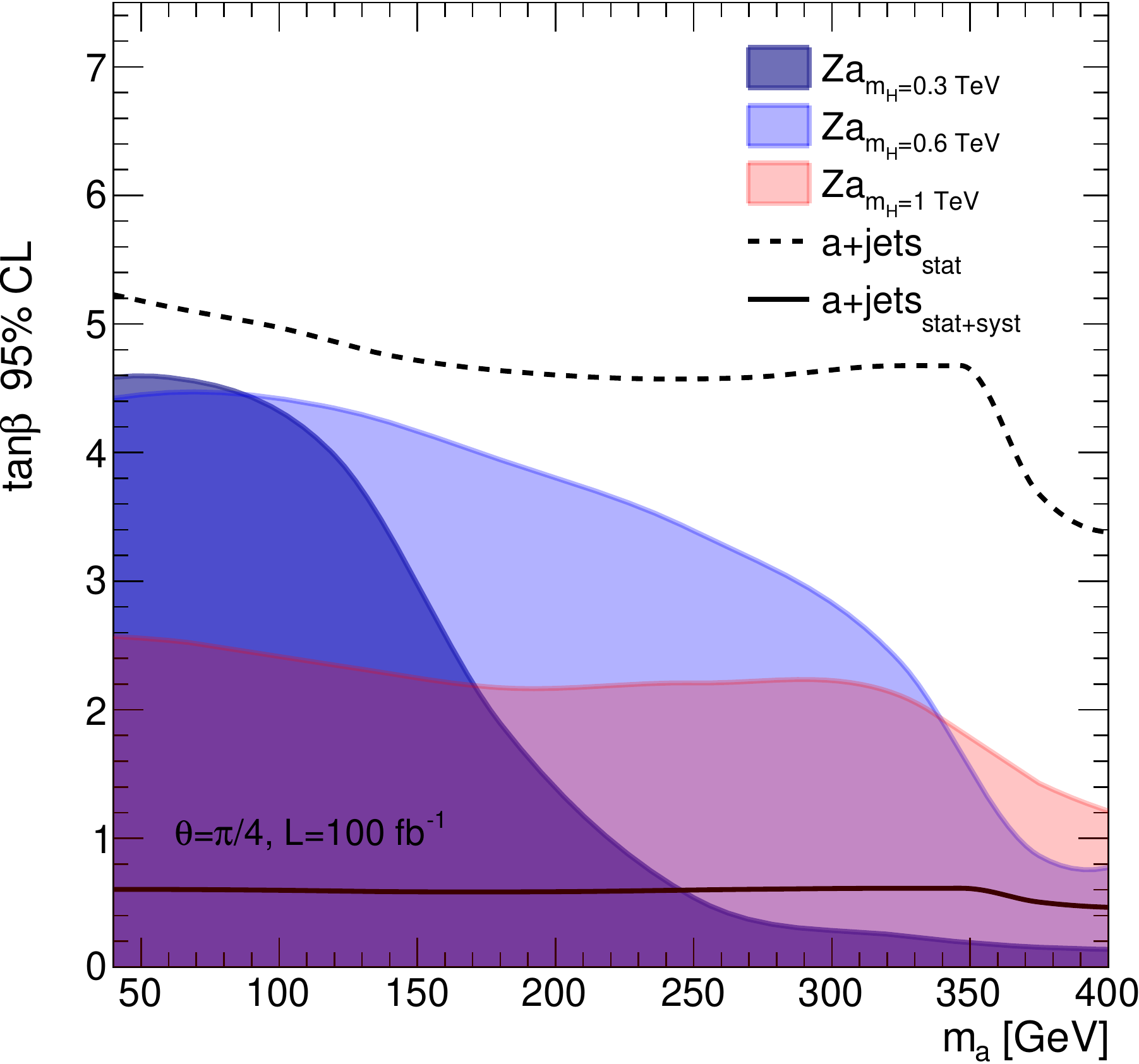}

\vspace{-2mm}

\caption{95\% CL bound on $\tan\beta$  as a function of the pseudoscalar mass $m_a$ for the 13~TeV LHC 
with $\mathcal{L}=100~\mathrm{fb}^{-1}$. We assume $\theta=\pi/4$ in the decoupling scenario ${m_A\gg m_a}$. 
The $a$+jets bound is displayed in two scenarios:  5\% systematic uncertainties on the background (black-full)
 and with only statistical uncertainties (black-dashed). The uncertainties are modelled as nuisance parameters.}
\label{fig:bound_ma} 
\end{figure}

In Fig.~\ref{fig:ptmiss1}, we show the $\slashed{E}_T$ distribution for
the signal with ${m_a=80}$~GeV. The sum of SM backgrounds was obtained
from~\cite{cms_monojet}, that accounts for $Z$+jets, $W$+jets,
$t\bar{t}$, dibosons $VV'$ and QCD multi-jet components.  We quantify
the signal sensitivity via a binned log-likelihood analysis to the
$\slashed{E}_T$ distribution, invoking the CL$_s$
method~\cite{Read:2002hq}. In Fig.~\ref{fig:bound_ma} we show the 95\%
C.L. bound on the $(m_a,t_\beta)$ plane for
$\mathcal{L}=100\,\mathrm{fb}^{-1}$. We stress however the strong
impact of systematic uncertainties on the mono-jet bounds: as shown in
Fig.~\ref{fig:bound_ma} (dashed-line), including the 5\% background
systematic uncertainty~\cite{Khachatryan:2014rra,Aaboud:2016tnv} weakens the mono-jet
sensitivity to $t_{\beta} \lesssim 0.6$, below the flavor bound for
2HDM Type I.

\vspace{2mm}

\begin{figure}[h!]
\centering
  \includegraphics[width=.4\textwidth]{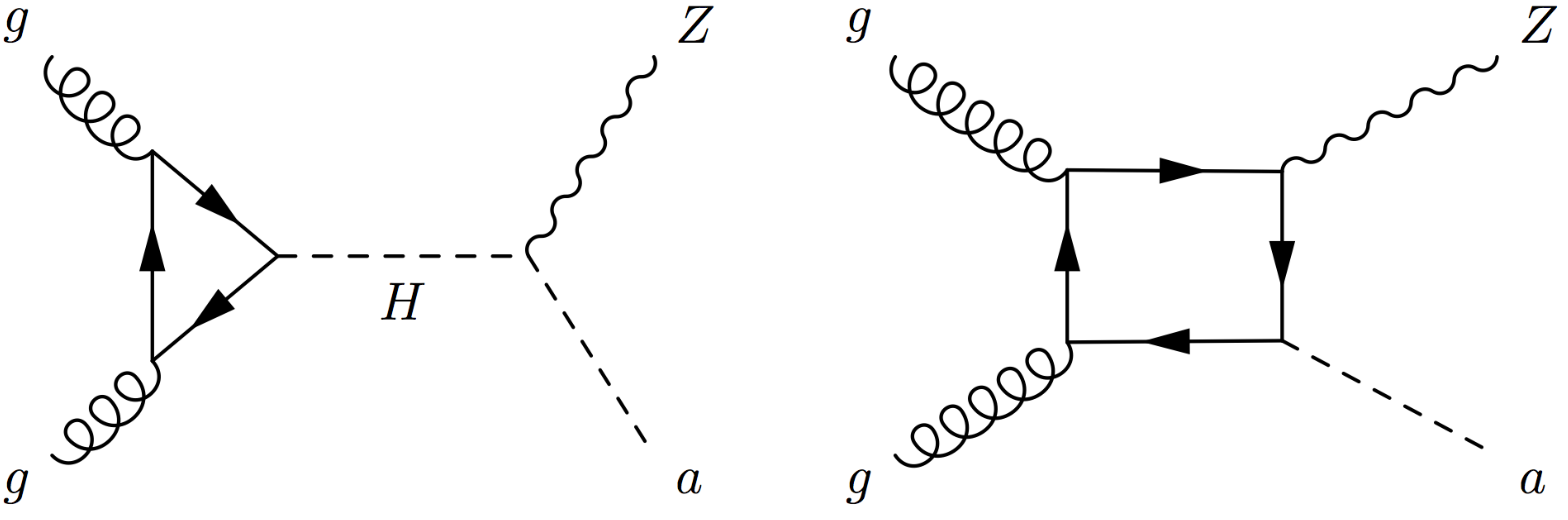}
  \vspace{-1mm}
  \caption{Sample of Feynman diagrams for the signal ${gg \rightarrow Za}$. }
\label{fig:feyn_az} 
\end{figure}

\noindent \textbf{LHC Phenomenology: Mono-$Z$ Searches}.
We now analyze the $p p  \to Z a$ channel. This channel can produce a very distinct collider 
signature characterised by boosted leptonic $Z$ decays recoiling against large amounts of missing 
energy from the  $a$ decays to Dark Matter $a\to \bar{\chi}\chi$~\cite{No:2015xqa,Goncalves:2016bkl}, see 
Fig.~\ref{fig:feyn_az}. The main backgrounds for this signature are top pair $\bar{t}t+$jets, diboson pair $V^{(*)}V'^{(*)}=WW,ZZ,WZ$ 
and $Z$+jets production.

We simulate our signal  and background samples with \textsc{Sherpa+OpenLoops}~\cite{sherpa,openloops,Goncalves:2015mfa,Goncalves:2016bkl}. 
The diboson and top pair samples are generated with the \textsc{MEPS@NLO} algorithm with up to one extra jet emission 
and the $Z$+jets with the same method merging up to two jets~\cite{mepsnlo}.  We also include the loop-induced gluon fusion 
contributions that arise for $ZZ$ and $WW$ production~\cite{Goncalves:2016bkl}. They are simulated at LO accuracy merged 
via the \textsc{CKKW} algorithm up to one  extra jet~\cite{ckkw}. Spin correlations and finite width effects from the vector bosons 
are accounted for in our simulation, as well as hadronisation and  underlying event effects~\cite{sherpa_spin}.
The pseudoscalar $a$ and heavy scalar $H$ widths are calculated from \textsc{Hdecay}~\cite{hdecay}.  

\begin{figure}[b!]
\centering
  \includegraphics[width=.48\textwidth]{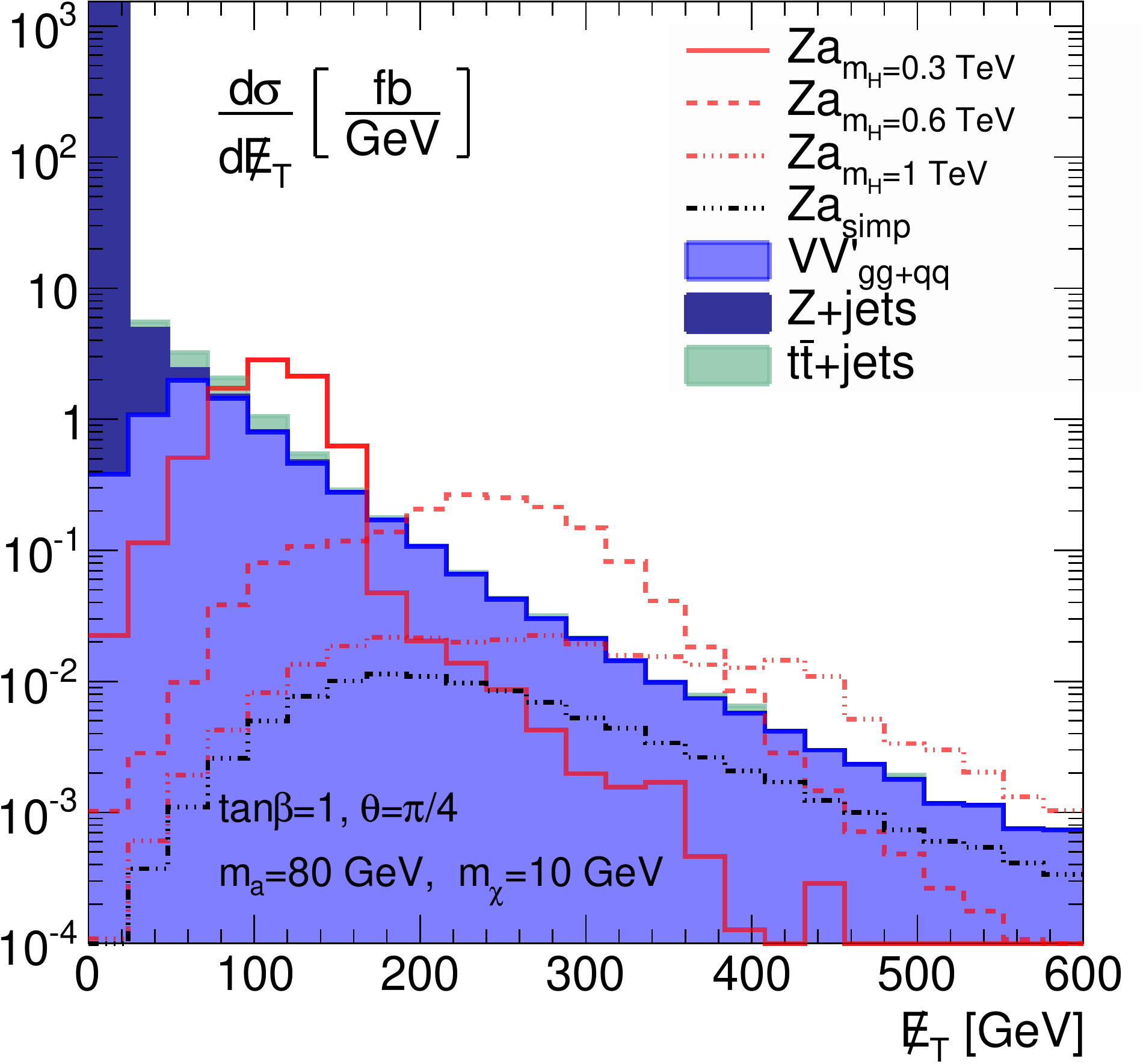}
  
  \vspace{-1mm}

  \caption{Signal (red) and background (blue/green) $\slashed{E}_T$
    distributions for $m_\chi=10$~GeV, $\mathrm{sin}^2(\theta)=1/2$,
    $\tan\beta=g_{\chi}=1$ and $m_a=80$~GeV (with $m_A \gg
    m_a$). Shaded (empty) histograms are (non-)stacked.
  We display the signal scenarios $m_{H} = 0.3,\,0.6\,,1$~TeV (red) and 
   within the simplified model framework (black).}
\label{fig:ptmiss2} 
\end{figure}

For the analysis, we require two same flavour opposite sign leptons with $p^{\ell}_{T}>20$~GeV, $|\eta_\ell|<2.5$ 
and ${|m_{\ell\ell}-m_Z|<15}$~GeV. As most of the sensitivity is in the boosted kinematics $\slashed{E}_T\gtrsim100$~GeV, 
where the $Z$ boson decay products are more collimated, we impose that $\Delta\phi_{ll}<1.7$. Jets are defined  via the anti-$k_T$ 
jet algorithm  $R=0.4$, $p_{Tj}>30$~GeV and $|\eta_j|<5$. To tame the $t\bar{t}$+jets background, we consider only the zero and
one-jet exclusive bins vetoing  extra jet emissions and b-tagged jets.  In Fig.~\ref{fig:ptmiss2},  the resulting $\slashed{E}_T$ 
distributions are shown, which highlights that for $\slashed{E}_T\gtrsim 90$~GeV,  the backgrounds $Z$+jets and $t\bar{t}$+jets get 
quickly depleted and the diboson $V V'$ becomes dominant.

Remarkably, for  $m_{H}>m_{a}+m_Z$, $H$ can be resonantly produced yielding a maximum in the $\slashed{E}_T$ 
spectrum~\cite{No:2015xqa}
\begin{equation}
\slashed{E}_T^{\mathrm{max}} \sim \frac{1}{2\,m_{H}}\sqrt{(m_{H}^2-m_{a}^2-m_Z^2)^2-4m_Z^2m_{a}^2}\,.
\label{eq:met}
\end{equation}
The position of the peak can then be shifted by changing $m_H$. In  Fig.~\ref{fig:ptmiss2} we show the $m_{H}=0.3,\,0.6,\,1$~TeV
scenarios, that  peak respectively at $\slashed{E}_T^{max}\sim 125,280,490$~GeV, following Eq.~\ref{eq:met}. Noticeably, the peak
gets less pronounced for larger $m_H$ due to the larger heavy resonance width $\Gamma_H$, smearing it out.

The 95\% C.L. signal sensitivity on the $(m_a, t_\beta)$ plane for the different $m_H$ benchmarks, through a 
two-dimensional ($\slashed{E}_T$ vs. number of jets ${n_j=0,1}$) binned log-likelihood, 
using the CLs method~\cite{Read:2002hq} with a 10\% systematic uncertainty on the
background rate~\cite{zh_cms}, is shown in Fig.~\ref{fig:bound_ma}
for $\mathcal{L}=100~\mathrm{fb}^{-1}$.

The results from Fig.~\ref{fig:bound_ma} stress that DM phenomenology at the LHC for the pseudoscalar portal is very different for
simplified model and consistent completion, particularly due to the presence of $H$ in the latter (and also $A$, $H^{\pm}$ if light),
which are required to be within LHC reach due to unitarity bounds. In this respect, mono-$Z$ searches yield
a significantly higher reach than mono-jet within the pseudoscalar portal, particularly for mono-jet 
background systematic uncertainties of order $5\%$ (as is the case in current experimental analyses~\cite{Khachatryan:2014rra,Aaboud:2016tnv}). 
Furthermore, for $m_{\chi} > m_a/2$, the mediator $a$ decays dominantly into SM particles (e.g. $a\to \bar{b}b$), and the 
process $p p \to H \to Z a$ also provides the leading probe of the mediator $a$~\cite{GMN,Dorsch:2016tab,Khachatryan:2016are}), 
complementing the associated pseudoscalar top channel $pp\rightarrow t\bar{t}a$~\cite{Goncalves:2016qhh,Casolino:2015cza} and 
significantly increasing the sensitivity of LHC searches to the parameter space region with $m_{\chi} > m_a/2$.

\begin{center}
 
\textbf{Summary}
 
\end{center}

In this Letter we have analyzed a minimal UV completion of the simplified pseudoscalar dark matter portal scenario.
In a minimal consistent setup, mixing between the light pseudoscalar and the new degrees of freedom (needed for
the existence of the portal) combined with unitarity of scattering amplitudes  require the new states to be around the TeV 
scale or below. This leads to key LHC phenomenology beyond the simplified model in the form of mono-$Z$ signatures, 
which yield a stronger sensitivity than the generic mono-jets analysis. Such outcome evinces the limitations of simplified 
models which are not  gauge-invariant, and evidences that the omission of degrees of freedom required for the theoretical 
consistency of simplified models can lead to a generic failure of these scenarios to capture the relevant physics.

\vspace{-1mm}

\begin{center}
\textbf{Acknowledgements} 
\end{center}

\vspace{-1mm}

\begin{acknowledgments}
We are grateful to the Mainz Institute for Theoretical Physics (MITP)
and the Universidade de S\~{a}o Paulo for the hospitality and partial
support. We also thank Tilman Plehn and Carlos Savoy for very useful
discussions, and Patrick Fox and Ayres Freitas for comments on the manuscript.  
The work of DG was funded by STFC  through  the IPPP grant and U.S. National Science 
Foundation under grant PHY-1519175. This project has received funding from 
the Centro de Excelencia Severo Ochoa No. SEV-2012-0249 (PM), and the following EU
grants: People Programe (FP7/2007-2013) grant No. PIEF-GA-2013-625809 EWBGLHC
(JMN); H2020 ERC Grant Agreement No. 648680 DARKHORIZONS (JMN); FP7 ITN INVISIBLES PITN-GA-2011-289442 (PM);
H2020-MSCA-ITN-2015/674896-Elusives (PM);
H2020-MSCA-2015-690575-InvisiblesPlus (PM). Fermilab is operated by
the Fermi Research Alliance, LLC under contract No. DE-AC02-07CH11359
with the United States Department of Energy.
\end{acknowledgments}

\end{document}